\documentclass{acm_proc_article-sp}

\begin{document}

\title{Software for Science: Some Personal Reflections\\
on Funding, Licensing, Publishing and Teaching\titlenote{This papers is licensed under a Creative Commons license to encourages sharing and remixing (with attribution). It may be combined with other ideas into the outcomes of the WSSSPE SC'13 workshop, if the author is acknowledged.}}
\subtitle{[4-page paper to be expanded and/or combined with others]
}

\numberofauthors{1}
\author{
\alignauthor 
Anne C. Elster \titlenote{I would also like to acknowledge all my wonderful graduate students, the students that take my courses as well as my other collaborators . My discussions with them give me continued inspiration, and have surely colored this work. However, the opinions expressed in this paper are my own.}\\
       \affaddr{Dept. of Computer and Information Science}\\
       \affaddr{Norwegian University of Science and Technology (NTNU)}\\
       \affaddr{Sem S\ae landsvei 9, N-7491 Trondheim, Norway}\\
       \email{elster@ntnu.no}
}

\date{Sept. 6, 2013}

\maketitle

\begin{abstract}

As computer hardware systems become more and more complex, software and tools lag more and more behind.
This is especially true for scientific software that often demands high performance, and thus needs to take advantage of parallelisms, memory hierarchies and other software and systems. How do we help bridge this ever-increasing gap?

This paper describes some of my experiences and thoughts regarding licensing, code sharing, code maintenance, open access publishing, and education and training. Details include my recent experiences with getting industrial funding for GPL licensed software, BSD licensing issues, sharing code on GitHub, and how I inspire students to take my 4th year Parallel Computing elective which this semester has over 50 students enrolled. Some thoughts and comments regarding why both optimization and data locality are such a central issue for scientific software is also included.

\end{abstract}


\section{Introduction}
 As stated in the description of WSSPE, the First Workshop on Sustainable Software for Science: Practice and Experience, to be held at SC'13\footnote{http://wssspe.researchcomputing.org.uk/}, progress in scientific research is dependent on the quality and accessibility of software at all levels. It is therefore critical to address many of the new challenges related to the development, deployment, and maintenance of reusable software. We also need to make sure scientists, researchers, and students acquire the new set of software-related skills and methodologies needed. NSF's {\it Vision and Strategy for Software}
that came out in February 2012 tries to highlight several of these issues and expects to encourage their solutions.  
The following sections address some of the issues this author has encountered and her thoughts related to them.

\section{Software funding}
As public funding in the US and Europe is generally declining, researchers and developer are becoming more and more dependent on industrial funding. Traditionally companies have been very protective of their IP. It is thus typical that software which originally was developed openly at first in research and university labs, later becomes proprietary as one sees potential for commercialization. 

\subsection{From MATLAB to Octave, R, etc.}
One clear example of this is MATLAB, a popular software package described by R. Schreiber
\footnote{(http://www.scholarpedia.org/article/MATLAB}. MATLAB was first developed by Cleve Moler in the early 1970's for his students to be able access the power of LINPACK and EISPACK without having to learn low-level FORTRAN programming. His line-based free MATLAB software with ASCII output and one data type (matrix double) grew quickly in popularity. {\it E.g.}, it became a central component in several of my graduate courses at Cornell in the 1980's, where I myself was introduced to it. I and many others have used it as a great prototyping environment for numerical algorithms.

MATLAB was re-written in C, added graphics ++, and commercialized through MathWorks, a privately held company. It has been very successful expanding the software with many coveted ``Toolboxes", including theDSP Systems Toolbox, statistics, etc, etc. It now also includes a Parallel Computing Toolbox that ``lets you solve computationally and data-intensive problems using multicore processors, GPUs, and computer clusters." 
\footnote{http://www.mathworks.se/products/parallel-computing/)} For more details on MATLAB history, see 
the Scholarpedia MATLAB article. 

Several open source projects with functionality similar to MATLAB, have now emerged. They include Octave\footnote{http://www.octave.org/} which many use, but is considered much slower, R\footnote{http://www.r-project.org} that focuses on Statistics, and NumPy and SciPy for Python enthusiasts.

\subsection{The price of software (and hardware)}

Few software projects have been as successful and/or costly as MATLAB. Its main licenses and toolboxes are known for being quite pricey for companies to buy as a development tool, but they do offer universities and students cheap and/or free 
(through donations) licenses for stripped-down versions. A similar scheme is, of course, used by almost all commercial vendors, including Microsoft (e.g. Visual Studio) and National Instruments. It is a clever way to get future industrial users 
"hooked" on their products". Similar strategies also exists for hardware donations. 

{\bf To reflect on:} Exposing the students to the software they will face in industry is clearly a good thing at some level, but does it make them too dependent on commercial software? I have had several master students so hooked on Windows and Visual Studio that they will not consider developing in or for Linux. What is this telling us? \\(They should take a look at Valgrind ...)

\subsection{Emerging funding schemes}
There are several new sources of software funding:

{\bf{App Stores}}\\
Mobile Applications (or apps) have through the vast public reach of Apple's and Google's on-line stores made several youngsters a good chunck of money despite low per-item fees. This happened to NTNU student H\aa kon Bertheussen who developed the game Wordfeud while he was still a student here. (As an aside, I can proudly report that he told me he made good use of the skills he developed in my parallel computing class.) However, it is unlikely that large scientific codes could gain similar statue as "`apps"' in the near term. On the other hand, there is already a vast number of tools for school kids that include simulations previously only seen in R \& D labs.

{\bf{Computer Games} }\\The gaming world's  demand for fast graphics has pushed the envelope of GPUs to the point where they are now very interesting platforms also for scientific codes. The games are also becoming more and more complex and include more and more realist physical simulations as complex as ``yesterday's" research science codes. Will the future bring a game with realistic weather simulations (maybe even tied into your Geolocation)? 

{\bf{Crowd funding}}\\Crowd funding schemes like Kickstarter.com, where the public is encouraged to fund development by pre-ordering products in exchange for early-access and swags. This has been particularly popular for some hardware vendors like Adapteva's Parallella project\footnote{http://www.kickstarter.com/projects/adapteva/\\parallella-a-supercomputer-for-everyone}
as well as mechanisms for some academics to provide on-line content as well as authors to write SW texts ({\it e.g.} iBook on SQL
\footnote{http://www.kickstarter.com/projects/522057582/sql-programming-the-easy-way-e-bookibook-and-cours?ref=live}
and PhP programming\footnote{http://www.kickstarter.com/projects/287976447/ \\
the-joy-of-php-programming-e-book-and-course?ref=live}).

{\bf{Embedded Ads}}\\
Another area that has really taken off is embedded advertizing, especially in browser-based products. Could this be an incentive for scientific software to develop browser-based GUIs? If so, how do we feel about this trend?

\section{SW Licensing and Patents}
Software is no longer either open on closed (only binaries available), but also licensed and patented. Even intentionally open codes are now typically licensed with an open source license to protect developers who want their codes to remain open, against liability, etc. Following is a brief discussion of three major licensing schemes as well as some reflexions regarding software patents.

\subsection{GPL Licenses}
The GNU General Public License (GNU GPL or GPL) was originally written by Richard Stallman, known as the father of the Free Software Foundation (FSF) for the GNU project\footnote{http://www.gnu.org/philosophy/pragmatic.html}. It has become one of the most popular free software licenses. It is a copy-left license, which is a general method for making a program (or other work) open, and requiring all modified and extended versions of the program to be open as well. GPL is thus used by developers that do not want their codes to be commercialized by others, and legally wants to force their codes to remain open. I.e., if you use a piece of GPL-licensed code in you own code, you have to also license your code under GPL. GPLv3 was released by FSF in June 2007 and is now the most used version. 

Several larger codes developed at universities and/or supported by larger companies, are now licensed under GPL. My personal experience is with Statoil who funds a PhD student that will work with DUNE and OPM, larger open code bases they have supported the development of (already licensed under GPL). It did take some work by the legal department of both my University and Statoil to agree to joint GPL licensing. Many university groups, including students like GPL, but many also rather use BSD.

\subsection{BSD Licenses}
The original BSD license was used for the Berkeley Software Distribution (BSD), a Unix-like operating system popular in the 1980's. BSD licenses impose minimal restrictions on the redistribution of covered software, i.e. unlike copy-left licenses, you may include BSD licensed code in commercial code (as well as in GPL code), but not {\it vice versa}. The original version has since been revised, and its descendants are now termed modified BSD licenses. The 2-clause BSD license a.k.a. Free\_BSD include a copyright retention for source code copies, and a note re. binary redistributions and disclaimer, whereas the newer 3-clause version  also includes the following statement: ``Neither the name of the <organization> nor the names of its contributors may be used to endorse or promote products
 derived from this software without specific prior written permission." 

Some Computer Science and Engineering academicians (including several people I know) do not want to touch GPL code as they find it too restrictive. They rather use BSD since they want their code to be "totally open" and leave room for the code to be reused in both proprietary codes as well as leave room for commercializing it later. 
Several software companies that donate  grants to universities wants the codes developed licensed under BSD for these reasons.

Both Apache License 2.0 and 3-clause BSD are considered compatible with version GPLv3. Apache 2.0 is not compatible with GPLv2 since it lacks certain patent termination and indemnification provisions.

\subsection{The Microsoft Reference License}
The Microsoft Reference License used by .NET, a Microsoft framework, specifies that the source code is made available for debugging purposes only, primarily to support integrated debugging of the BCL in Visual Studio. As mentioned, several students (and corporations) use .NET and Visual Studio, but there are research groups, including my own, that prefer to shun .NET due to the restrictions of this license.

\subsection{Software Patents and Uniqueness}
Most countries place some limits on software patents, which are not legally defined. In Europe, ``computer programs as such"are excluded from being patented, and European Patent Office policy is consequently that a computer program is not patentable if it does not have the potential to cause a ``further technical effect" beyond the inherent technical interactions between hardware and software. U.S. patent law excludes ``abstract ideas", which has been used to refuse some software patents. 

Back in early 1987 when I presented my linear bit-reversal algorithm \cite{5161105}, I was told by the IBM people presentthat I should have patented it. To me that seemed strange, since I fiercely believed in open source and code sharing.

\section{Publishing}
There is still the challenge of making ones software ideas known. I published my bit-reversal algorithm at ICASSP in 1989 \cite{266624} since I felt that the signal processing audience would best appreciate it. It was known in 1996 to still be the fastest \cite{karp1996}. However, I have discovered several papers later in IEEE journals and elsewhere that pretty much published the same algorithm. (I also stopped a couple as a reviewer.) The powerful search engines of today should make things easier, but I have a feeling a lot of publications are not as unique as claimed. Will open access publishing help?

{\bf Open Access} is now promoted by NIH \footnote{http://publicaccess.nih.gov/policy.htm} and other government agencies, several of whom now demand that works resulting from publicly funded projects be published open access. Several open access publications, some serious, some not have thus emerged. Some traditional journals offer authors open access of their articles for a fee. E-print services that enable self-publishing, such as arXiv.org where this paper is published, are also becoming popular. {\it IEEE Access} is a recent interesting addition which provides an arena for inter-disciplinary work on-line. We recently got a paper accepted with this journal \cite{falch-elster-2013} and were able to include movies associated with our article.  A rising concern is the non-trivial fee such journals ask for up front, which can  be a hindrance to some. 

Another downside is that open access journals do not yet have the clout of other publication channels. In fact, in recent discussions among women, many feel that young non-established researchers, especially women and other minorities that do not yet have tenure, should not risk their careers by publishing in ``lesser" journals. We are told we should first published in the most prestigious (often for our libraries the most expensive) journals in order to get respect in our community and be thus well set up for promotion. 

Most, if not all traditional and open access computing journals {\bf rarely considers the reproducibility, quality, or even availability of the software their articles may be built on!} This both hinders code re-use and tarnishes the reputation of our field versus other field where reproducibility is a must.  

Another challenge for computing systems is that often our conferences may have much lower acceptance rates that most journals in the field, whereas in other fields one may only publish abstracts at conferences, and thus consider anything but journal papers``light weight". Much more can be said about the whole``publish or perish" tenure system world-wide that encourages LPUs (least publishable units), but that will digress a bit from the core of this paper.

We also seem to be moving away from providing longer versions of our work as Technical Reports. This is a negative trend. On-line publishing is becoming easier and more feasible, but I fear that as long as one does not get full credit for technical reports as "real" publications, the pressures of LPUs tarnishes this great open venue for documenting algorithms and methods better so that they can get reused.

\section{Code reuse}
Unfortunately, maintaining codes is always a challenge. Hardware changes often, patches and updates to compilers, libraries and services are a constant. It really takes some dedicated people to keep it all up. One strategy is to release the code to a company or institution large enough to have resources to maintain and develop the codes. This then brings us back to many issues related to funding and licensing. How do we ease the pain of code maintenance while still ensuring great progress?

A solution would be to start requiring all published papers (at least in respected fora) to also provide open source versions of the software on which they are based. My group has started doing this using GitHub rather than through our own wiki-based server. It would be preferrable if the code could be stored online with the paper where it is published.

{\bf Software libraries} are a great mechanism to enable code reuse of commonly used routines and/or highly optimized kernel functions. Care must be taken, as we did for MPI\footnote{http://www.mpi-forum.org/}, so that programs can access library functions both directly as well as through other libraries or linked code using the library. One of the largest impediments to progress in this area is not enough application developers are aware of the great libraries out there.
Many hours are thus wasted by our community reimplementing codes have already been implemented in great libraries. 

Similarly, {\bf reproducibility} is a huge problem in our field, and it is not just because of round-off when running large codes. Even more efforts are spent re-implementing other researchers' and developers' codes that have only vaguely been described in published papers.

\section{Education and training}

I had over 50 students in both my compiler course last spring and my parallel computing class this fall (2013). I believe a large reason why these elective courses have become so popular is a) these students see their relevance and b) they really like the programming assignments.

I often tell my students and audiences that if you want your code to be performant, is it like in real estate; it is all about location, location and location of your data. This is true whether you worry about fitting all your data into RAM (and avoiding thrashing the disk (or other external devices)) or fitting the data into various levels of cache. As soon as you run out of registers, you have to worry about data location. Of course, this becomes a further worry when there are multiple ALUs and processor cores to feed, and as soon as you add other devices, like GPUs , with their own memory banks, the problems just gets even more challenging. 

{\bf Locality:} To quote what is said about data locality in the NSF/TCPP document (appendix):
``The performance advantages of locality are easy to explain and can be illustrated by taking examples from a wide spectrum of data access scenarios. This includes cache data locality in the programming context, memory locality in paging context, disk access locality, locality in the context of virtualization and cloud computing, etc. Both spatial and temporal aspects of locality must be clarified by illustrating situations where only one kind or both may be present. Simple eviction/prefetching policies to take advantage of locality should also be illustrated with examples. Relationship of temporal locality to the notion of working set should also be explained." 

A chapter I have proposed for the NSF/TCPP CDER book project 
will address several of the above related locality and/or performance topics. 
The material we develop will be made freely available. One of the goals is to have PDC (Paralell and Distributed Computing)topics introduced much earlier in the curriculum. 

{\bf Starting with MPI:} Unlike several of my colleagues I like to start my parallel programming students on MPI and serial code optimization (for cache) before OpenMP, Pthreads and Stream-based programming in CUDA and/or OpenCL since it forces them to think about locality. This has been very successful, and Here is how it may be approached: 

To get students started on thinking about data locality and synchronization, 
I use the analogy of e-mail sends and receives.
I initially limit the MPI intro to the 6 basic functions of init, rank, size, send, receive and finalize, and illustrate a parallel "Hello, world" + hello with data.   I ask students how they share data like photos and film-clips. Often, it is through mailing or MMS-ing, but it could be through Dropbox and GoogleDocs, which implies synchronization.
After introducing barriers and broadcasts, we often include problems sets that involve solving a PDE using Poisson's Equations, or other applications that need border exchanges.

I then jump to optimization and caching, including how auto-tuning libraries like BLAS and FFTW \cite{1386650} are so hard to beat since they address cache locality. I gloss over OpenMP (from a location point of view) and dive into streaming and data locality for GPU programming -- where again locality becomes a central issue \cite{parco09-mulit-gpu, aqrawi-elster-2011}.

\section{Conclusions} 

The challenges of how to develop and maintain sustainable scientific software remain.
This paper addressed some of the central issues regarding this including  thoughts on funding, licensing, publishing and training.

Hopefully, some of the experiences and thought presented here will feed into a collaborative writing of one or more journal publications on this topic. Only a few references are included, most related to the authors own experiences, due to the 4-page WSSSPE limit for this version.

\subsection*{Acknowledgments}
In addition to her graduate students and others mentioned earlier, the author would like to especially thank Statoil, and NTNU for their support of her student and HPC-lab.

\bibliographystyle{abbrv}
\bibliography{sw-ref}  

\begin{thebibliography}{1}

\bibitem{aqrawi-elster-2011}
A.~A. Aqrawi and A.~C. Elster.
\newblock Bandwidth reduction through multithreaded compression of seismic
  images.
\newblock In {\em Parallel and Distributed Processing Workshops and Phd Forum
  (IPDPSW), 2011 IEEE International Symposium on}, pages 1730--1739. IEEE,
  2011.

\bibitem{266624}
A.~C. Elster.
\newblock Fast bit-reversal algorithms.
\newblock In {\em Acoustics, Speech, and Signal Processing, 1989. ICASSP-89.,
  1989 International Conference on}, pages 1099--1102 vol.2, 1989.

\bibitem{5161105}
A.~C. Elster and J.~C. Meyer.
\newblock A super-efficient adaptable bit-reversal algorithm for multithreaded
  architectures.
\newblock In {\em Parallel Distributed Processing, 2009. IPDPS 2009. IEEE
  International Symposium on}, pages 1--8, 2009.

\bibitem{falch-elster-2013}
T.~L. Falch, J.~Fl{\o}ystad, D.~W. Breidby, and A.~C. Elster.
\newblock Gpu-accelerated visualization of scattered point data.
\newblock {\em Access, IEEE}, 1:0--0, 2013.

\bibitem{1386650}
M.~Frigo and S.~Johnson.
\newblock The design and implementation of fftw3.
\newblock {\em Proceedings of the IEEE}, 93(2):216--231, 2005.

\bibitem{karp1996}
A.~H. Karp.
\newblock Bit reversal on uniprcoessors.
\newblock {\em SIAM Review}, 938(1):1--26, 1996.

\bibitem{parco09-mulit-gpu}
D.~G. Spampinato, A.~C. Elster, and T.~Natvig.
\newblock Modelling multi-gpu systems.
\newblock In {\em Advances in Parallel Computing ( PARCO 2009)}, pages
  562--569. IOP, 2009.

\end{thebibliography}

\balancecolumns
\end{document}